\documentclass[fleqn]{elsart}
\usepackage{latexsym,amssymb,amsmath,amsfonts,epsfig}

\makeatletter \@addtoreset{equation}{section} \makeatother
\renewcommand{\theequation}{\thesection.\arabic{equation}}
\renewcommand{\today}{December 21, 2004}
  
\newcommand{\no}{\nonumber}
\newcommand{\bR}{\mathbb{R}}
\newcommand{\bC}{\mathbb{C}}
\newcommand{\bZ}{\mathbb{Z}}

\def\e{{\rm e}}

\newcommand{\frachalf}{{\textstyle \frac{1}{2}}}

\renewcommand{\i}{{\rm i}}

\newcommand{\tfr}[2]{{\textstyle \frac{#1}{#2}}}

\newcommand{\pdq}[2]{\frac{\partial #1}{\partial #2}}
\newcommand{\dx}{\!{\rm d}^4x\,\,}
\newcommand{\dvx}{\!{\rm d}^3 x\,\,}

\newcommand{\dint}[1]{\!{\rm d}#1\,\,}

\newcommand{\tr}{\text{tr}}

\newcommand{\vect}     [1]{\left( \begin{array}{c} #1 \end{array} \right)}
\newcommand{\vectleft} [1]{\left( \begin{array}{l} #1 \end{array} \right)}
\newcommand{\twomat}   [1]{\left( \begin{array}{cc} #1 \end{array} \right)}
\newcommand{\threemat} [1]{\left( \begin{array}{ccc} #1 \end{array} \right)}
\newcommand{\threematw}[1]{\left( \begin{array}{c@{\hspace{4mm}}c@{\hspace{4mm}}c} #1 \end{array} \right)}

\newcommand{\sixmat}   [1]{\left( \begin{array}{cccccc} #1 \end{array} \right)}

\newcommand{\pr}{\partial}

\newcommand{\id}{\mathbf{1}}

\newcommand {\rhs} {right-hand side}

\newcommand{\bcs}  {boundary conditions}
\newcommand {\ncs} {noncontractible sphere}
\newcommand {\ncl} {noncontractible loop}

\newcommand{\YM}   {Yang--Mills}
\newcommand{\YMth} {Yang--Mills theory}
\newcommand{\YMH}  {Yang--Mills--Higgs}
\newcommand{\YMHth}{Yang--Mills--Higgs theory}
\newcommand{\SM}   {Standard Model}

\begin{document}
\noindent  Nucl. Phys. B 709 (2005) 171 
\hfill     hep-th/0410195,    %%hep-th/0410195
           KA--TP--08--2004\vspace*{0.5cm}\newline
\runauthor{F.R. Klinkhamer, C. Rupp}
\begin{frontmatter}
\title{A sphaleron for the non-Abelian anomaly} 
\author{F. R. Klinkhamer},
\ead{frans.klinkhamer@physik.uni-karlsruhe.de}
\author{C. Rupp}
\ead{cr@particle.uni-karlsruhe.de}
\address{Institute for Theoretical Physics,      
University of Karlsruhe (TH), 76128 Karlsruhe, Germany}
\begin{abstract}
A self-consistent \emph{Ansatz} for a new sphaleron of $SU(3)$
Yang--Mills--Higgs theory is 
presented. With a single triplet of Weyl fermions added, there 
exists, most likely, 
one pair of fermion zero modes, which is known to
give rise to the non-Abelian (Bardeen) anomaly as a Berry phase.
The corresponding $SU(3)$ gauge field 
configuration could take part
in the nonperturbative dynamics of  Quantum Chromodynamics.
\end{abstract}
\end{frontmatter}

\section{Introduction}
\label{sec:Introduction}

The Hamiltonian formalism makes clear that
chiral anomalies \cite{A69,BJ69,B69,WZ71,tH76prl,W82,AG84} are directly
related to particle production \cite{NA85}.
The particle production, in turn, traces back to a point
in configuration space (here, the mathematical space of
three-dimensional bosonic field configurations with finite energy) 
for which the massless  Dirac Hamiltonian has zero modes. 
This point in configuration space may correspond to a so-called
sphaleron (a static, but unstable, classical solution), provided appropriate
Higgs fields are added to the theory in order to
counterbalance the gauge field repulsion.

The connection of the triangle (Adler--Bell--Jackiw) anomaly \cite{A69,BJ69} 
and the sphaleron $\mathrm{S}$ \cite{DHN74,M83,KM84} is well-known. 
An example of an anomalous but consistent theory is the chiral 
$SU(2)\times U(1)$ gauge theory of the electroweak \SM, where the anomaly 
appears in the divergence of a 
non-gauged vector current (giving rise to $B+L$ 
nonconservation \cite{tH76prl}, with $B$ and $L$ the baryon and lepton number).  
The electroweak sphaleron $\mathrm{S}$ \cite{KM84,KKB92} is believed 
to contribute to fermion-number-violating processes in the early universe.

\begin{figure}[t]
\begin{center}
\includegraphics[width=10cm]{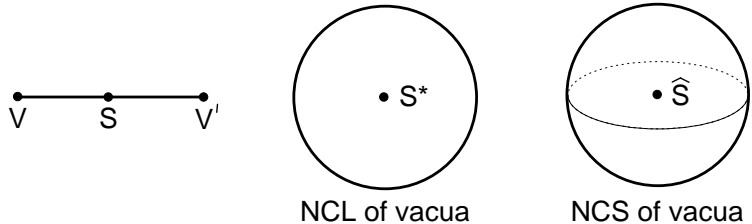}
\end{center}
\caption{Configuration space: vacua and sphalerons.}
\label{FIG-SSstarSbar} 
\end{figure}

The point $\mathrm{S}$ in configuration space locates
the energy barrier between vacua V and V$^\prime$, 
where V$^\prime$ is the gauge transform of V with a gauge function 
corresponding to the generator of the homotopy group 
$\pi_3[SU(2)]$. 
The sphaleron $\mathrm{S}$ 
has one fermion zero mode for each isodoublet of left-handed Weyl fermions
\cite{N75,KM84aps,BK85,R88,KB93}. 
A one-dimensional slice of configuration space  
is sketched in the left part of Fig.~\ref{FIG-SSstarSbar}. 

The nonperturbative $Sp(n)$ (Witten) anomaly \cite{W82} is also 
related to a sphaleron, namely the solution $\mathrm{S}^\star$ 
constructed in Ref.~\cite{K93}. 
The electroweak sphaleron $\mathrm{S}^\star$ sets the energy barrier 
for the Witten anomaly (which cancels out in the \SM)
and may play a role for the asymptotics of perturbation theory \cite{KW96}
and in multiparticle dynamics \cite{K98zm}.

An example of an anomalous and inconsistent theory is the chiral 
$Sp(2)=SU(2)$ gauge theory with a
single doublet of left-handed Weyl fermions.
In this case, there is a
\ncl~(NCL) of gauge field vacua 
(corresponding to the nontrivial element of 
$\pi_4[SU(2)]$)
at the edge of a two-dimensional disc in configuration space. 
For one ``point'' of the disc ($\mathrm{S}^\star$ in the \YMHth), there are
crossing fermionic levels.  The sphaleron $\mathrm{S}^\star$ 
has indeed two fermion zero modes \cite{K98zm,KR03}. 
The Witten anomaly results from the Berry phase factor $-1$ 
for a single closed path around these
degenerate fermionic levels \cite{B84}. This 
phase factor gives a M\"{o}bius bundle structure over the edge of the 
disc (the NCL of gauge field vacua),
which prevents the continuous implementation of Gauss' law.
In other words, the nontrivial
Berry phase factor presents an insurmountable obstruction to the 
implementation of Gauss' law 
and physical states are no longer gauge invariant \cite{W82,NA85}. 
A two-dimensional slice of $SU(2)$ \YMH~configuration space  
is sketched in the middle part of Fig.~\ref{FIG-SSstarSbar}.

There remains the perturbative
non-Abelian (Bardeen) anomaly \cite{B69,WZ71,AG84}. An example of an
anomalous and inconsistent theory is the chiral $SU(3)$ gauge theory with a
single triplet of left-handed Weyl fermions.
Now, there is a \ncs~(NCS) of gauge field vacua 
(corresponding to the generator of 
$\pi_5[SU(3)]$)
at the border of a three--ball in configuration space. 
By the family index theorem, there are 
crossing energy levels of the Dirac Hamiltonian
at one ``point'' of the three--ball \cite{NA85}. 
The non-Abelian anomaly is the Berry phase \cite{B84} from these
degenerate fermionic levels. 

\begin{figure}
\begin{center}
\includegraphics[width=4cm]{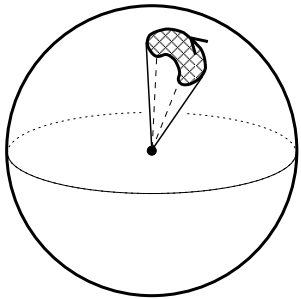}
\end{center}
\caption{$SU(3)$ Bardeen anomaly as Berry phase.}
\label{FIG-NCSbardeen}
\end{figure}

Again, a nonzero Berry phase ($\mathrm{mod}\: 2\pi$) presents an 
obstruction to the implementation of Gauss' law. 
The Berry phase of a loop on the NCS 
considered does not vanish and is, in fact, 
equal to one half of the solid angle subtended 
at the ``point'' with crossing energy levels by the loop; 
see Refs.~\cite{NA85,B84} for further details and
Fig.~\ref{FIG-NCSbardeen} for a sketch. 
It has been conjectured \cite{K98} that 
this special ``point'' in configuration space corresponds to a new type of
sphaleron in $SU(3)$ \YMHth~(denoted 
by $\widehat{\mathrm{S}}$ in the right part of Fig.~\ref{FIG-SSstarSbar})  
and the goal of the present article is to establish the 
basic properties of this sphaleron.

The outline of our article is as follows.  
Section~\ref{sec:Theory} recalls the action of chiral $SU(3)$ gauge 
theory, mainly in order to establish notation.
Section~\ref{sec:Topology} discusses the topology 
responsible for the new sphaleron, together with approximate 
bosonic field configurations.
Section~\ref{sec:SphaleronAnsatz} presents the generalized \emph{Ansatz} 
for the bosonic fields, which can be shown to solve the field
equations consistently.
Section~\ref{sec:Fermionzeromodes} gives the corresponding \emph{Ansatz} 
for the fermionic field, together with analytic and numerical 
results which suggest the existence of two fermion zero modes.
Section~\ref{sec:Conclusion} presents some concluding remarks.
Three appendices give 
the \emph{Ansatz} energy density for the bosonic fields, 
an argument in favor of a nontrivial regular solution 
of the reduced field equations,
and the fermion zero modes for a related gauge field background.

\section{Chiral $SU(3)$ gauge theory}
\label{sec:Theory}

In this article,  we consider a relatively simple theory, namely
$SU(3)$ \YMHth~with a single triplet of complex scalar fields and 
a single triplet of massless left-handed Weyl fermions, both in the 
complex  $\mathbf{3}$ representation of $SU(3)$. 
Our results can be readily extended to other gauge groups $G$ with
$\pi_5[G]= \Zset$ 
and appropriate representations.\footnote{Non-Abelian anomalies   
also appear in certain $(2n)$-dimensional chiral gauge theories
with finite 
$\pi_{2n+1}[G]$.
The requirement 
$\pi_{2n+1}[G]= \Zset$ 
is only a \emph{sufficient} condition
for having non-Abelian anomalies in $2n$ spacetime dimensions;
cf. Section 5 of Ref.~\cite{AG84}.}

The chiral $SU(3)$ gauge theory has the following classical action: 
\begin{equation} 
S = \int_{\bR^4}\; \dx
\left\{ \frachalf \, \tr\, F_{\mu\nu}  F^{\mu\nu} +
       (D_\mu  \Phi)^\dagger \,(D^\mu \Phi)     -
       \lambda \, \left( \Phi^\dagger \Phi - \eta^2 \right)^2
+\i \, \bar \psi\, \sigma^\mu D_\mu\, \psi
 \right\}\,,
\label{actionYMHW}
\end{equation}
where $F_{\mu\nu}\equiv\partial_\mu A_\nu -\partial_\nu A_\mu + g [A_\mu,
A_\nu]$ is the  $SU(3)$  Yang--Mills field strength tensor
and $D_\mu \equiv (\partial_\mu + g A_\mu)$
the covariant derivative for the $\mathbf{3}$ representation.

The $SU(3)$ Yang--Mills gauge field is defined as 
\begin{equation}
A_\mu (x) \equiv A_\mu^a (x)\, \lambda_a / (2\i)\,, 
\label{Adef}
\end{equation}
in terms of the eight Gell-Mann matrices  
\begin{align}
\lambda_1 &= 
\threematw{0 & 1 & 0 \\[-1ex] 1 & 0 & 0 \\[-1ex] 0 & 0 & 0 }\,, & 
\lambda_2 &= 
\threematw{0 & -\i & 0 \\[-1ex] \i & 0 & 0 \\[-1ex] 0 & 0 & 0 }\,, & 
\lambda_3 &= 
\threematw{1 & 0 & 0 \\[-1ex] 0 & -1 & 0 \\[-1ex] 0 & 0 & 0 }\,,
\nonumber\\[1em] 
\lambda_4 &= 
\threematw{0 & 0 & 1 \\[-1ex] 0 & 0 & 0 \\[-1ex] 1 & 0 & 0 }\,, &
\lambda_5 &= 
\threematw{0 & 0 & -\i \\[-1ex] 0 & 0 & 0 \\[-1ex] \i & 0 & 0 }\,,&
\lambda_6 &= 
\threematw{0 & 0 & 0 \\[-1ex] 0 & 0 & 1 \\[-1ex] 0 & 1 & 0 }\,, &
\nonumber \\[1em]
\lambda_7 &= 
\threematw{ 0 & 0 & 0 \\[-1ex] 0 & 0 & -\i \\[-1ex] 0 & \i & 0}\,,&
\lambda_8 &=  \frac{1}{\sqrt{3}}
\threematw{1 & 0 & 0 \\[-1ex] 0 & 1 & 0 \\[-1ex] 0  & 0 & -2 }\,.
\end{align}
The field $\Phi(x)$ is a triplet of complex scalar fields, which 
acquires a vacuum expectation value $\eta$ due to the Higgs potential term
in the action (\ref{actionYMHW}).   
The field $\psi(x)$ is a triplet of
two-component Weyl spinors and $\bar\psi(x)$ its complex
conjugate. Also, we define $(\sigma^\mu) \equiv (\id, -\vec \sigma)$ 
in terms of the Pauli matrices $\vec \sigma$ and
use the Minkowski space metric with 
$g_{\mu\nu}(x)={\rm diag}(+1,-1,-1,-1)$
and natural units with $\hbar=c=1$.

For static three-dimensional field configurations, we generally employ 
the spherical polar coordinates $(r,\theta,\phi)$ but sometimes also
the cylindrical coordinates $(\rho,\phi,z)$, with
$\rho \equiv r\,\sin\theta$ and $z\equiv r\,\cos\theta$.

\section{Topology and approximate sphaleron} 
\label{sec:Topology}

The basic motivation for our search of a new sphaleron 
($\widehat{\mathrm{S}}$) in $SU(3)$ \YMHth~has been given in the penultimate 
paragraph of Section~\ref{sec:Introduction}
and was summarized by Fig.~\ref{FIG-NCSbardeen}. 
We start with the bosonic fields of the theory (\ref{actionYMHW})
and introduce a  topologically nontrivial map $S^3 \wedge S^2 \to SU(3)$,
with  parameters $(\psi,\mu,\alpha)$ and 
coordinates $(\theta,\phi)$ on the ``sphere at infinity.''
Here, $\wedge$ denotes the so-called smash product which topologically
transforms the Cartesian product $S^p \times S^m$ to $S^{p+m}$ by
contracting, for fixed points $x_0\in S^p$ and $y_0\in S^m$, the set
$\{x_0\}\times S^m \cup S^p \times \{ y_0\}$ to a single point. 
Topologically, one has therefore $S^p \wedge S^m \sim S^{p+m}$; 
cf. Ref.~\cite{KR03}.

The procedure to be followed
is similar to the one used for the sphalerons S \cite{M83,KM84}
and $\mathrm{S}^\star$ \cite{K93}, which employed 
maps  $S^1 \wedge S^2 \to SU(2)$
and   $S^2 \wedge S^2 \to SU(2)$, respectively.
The reason behind this particular choice of smash products is that 
the search of new solutions is best performed in a fixed gauge, 
which we take to be the radial gauge
(hence, $\theta$ and $\phi$ appear in the maps, but not $r$). 
The three--ball of Fig.~\ref{FIG-NCSbardeen} or the right part of 
Fig.~\ref{FIG-SSstarSbar} becomes in the radial gauge a three--sphere 
(parametrized by $\psi$, $\mu$, and $\alpha$)
with $\widehat{\mathrm{S}}$ at the top and $\mathrm{V}$  
at the bottom, where the ``height'' corresponds to the energy of the
configuration. See, e.g., Refs.~\cite{M83,KR03} for further details.

The following ``strikingly simple''  map \cite{PR03} is a generator of 
$\pi_5[SU(3)]=\bZ$:
\begin{equation}
U: S^5 \to SU(3)\,, \quad
U(z_1,z_2,z_3) = \threemat{
z_1^{2}              &\;\; z_1\,z_2- \bar z_3     &\;\; z_1\, z_3 +\bar z_2 \\
z_1\,z_2+ \bar z_3   &\;\; z_2^2                  &\;\;  z_2\, z_3- \bar z_1 \\
z_1\, z_3 - \bar z_2 &\;\;  z_2 \, z_3 + \bar z_1 &\;\; z_3^{2}
}\,, \label{Udef}
\end{equation}
with $z_1,z_2,z_3 \in \bC$ and $|z_1|^2+ |z_2|^2+ |z_3|^2 =1$.
An appropriate parametrization for our purpose is
\begin{equation}
\vect{z_1\\[1ex]z_2\\[1ex]z_3} \equiv 
\vect{ 1-\cos^2\frac{\theta}{2} \, (1-\cos\psi) 
+\i \, \cos\frac{\theta}{2} \sin\psi \cos\mu\\[1ex]
\e^{\i\phi}\,
  \sin\frac{\theta}{2}\,\cos\frac{\theta}{2} \, (1-\cos\psi)
\\[1ex]
\cos\frac{\theta}{2}\, \sin\psi\, \sin\mu\, (\sin\alpha 
+ \i\, \cos\alpha) } \,,
\label{smashparam}
\end{equation}
with angles $\psi,\mu,\theta \in [0,\pi]$ 
and  $\alpha,\phi \in [0,2\pi]$.
At first sight, Eq.~(\ref{smashparam}) 
constitutes a map $S^3 \times S^2\to S^5$, with
$S^3$ parametrized by spherical polar coordinates $(\psi, \mu, \alpha)$ 
and  $S^2$ by $(\theta, \phi)$. 
But the expression on the \rhs\ of
Eq.~(\ref{smashparam}) corresponds to the single
point $(z_1, z_2, z_3)=(1,0,0)$ 
if either $\psi=0$ or $\theta=\pi$ and the map (\ref{smashparam})
is effectively  $S^3 \wedge S^2 \to S^5$. 
The fixed points $x_0 \in S^3$ and $y_0\in S^2$ needed for 
the smash product are thus given by
$\psi=0$ and $\theta=\pi$, respectively.

For $\psi=0$, the matrix $U(z_1,z_2,z_3)$ is independent of 
the parameters $\mu$ and $\alpha$ and the coordinates
$\theta$ and $\phi$,
\begin{equation}
V \equiv  U(\psi,\mu,\alpha,\theta,\phi)\bigr|_{\,\psi=0} =  
\threematw{ 
1
& 0
& 0  \\[-1ex]
0 
& 0
& -1  \\[-1ex]
0
& +1
& 0 }\,.
\label{Vdef}
\end{equation}
For $\psi=\pi$, the matrix $U(z_1,z_2,z_3)$ is  independent of $\mu$ 
and $\alpha$ but does depend on $\theta$ and $\phi$,
\begin{align}   
W(\theta,\phi) &\equiv  
U(\psi,\mu,\alpha,\theta,\phi)\bigr|_{\,\psi=\pi} \no\\[2mm]
&=  \threemat{ 
\cos^2\theta 
&\;\; -\cos\theta\, \sin\theta\, \e^{\i\phi} 
&\;\;\sin\theta \,\e^{-\i\phi}   \\[1ex]
-\cos\theta\, \sin\theta\, \e^{\i\phi} 
&\;\; \sin^2\theta \, \e^{2\i\phi}
&\;\; \cos\theta  \\[1ex]
-\sin\theta\, \e^{-\i\phi} 
&\;\; -\cos\theta 
&\;\; 0 }\,.
\label{Wdef}
\end{align}
These two matrices have the following rotation and 
reflection symmetries: 
\begin{subequations} \label{symmetriesVW} 
\begin{eqnarray}
&&
\partial_\phi\, M +
\frac{\i}{2}\, \left( \lambda_3-\sqrt{3}\,\lambda_8\right)  M 
+\frac{\i}{2}\, M  \left( \lambda_3-\sqrt{3}\,\lambda_8 \right)
=0 \,, 
\label{axialsymmVW}\\[2mm]
&&
V(\theta,\phi) = V(\pi-\theta,\phi)\,,
\label{reflsymV}\\[2mm]
&&
\threematw{-1 &0&0\\[-1ex] 0&1&0\\[-1ex] 0&0&-1}\, W(\theta,\phi)\, 
\threematw{-1 &0&0\\[-1ex] 0&1&0\\[-1ex] 0&0&-1} = 
W(\pi-\theta,\phi)\,,
\label{reflsymW}
\end{eqnarray}
\end{subequations}
for $M=V$ and $W$ as defined by Eqs.~(\ref{Vdef}) and (\ref{Wdef}),
with a trivial $(\theta,\phi)$ dependence for $V$.  

Observe that the general 
matrix $U(\psi,\mu,\alpha,\theta,\phi)$ possesses the reflection
symmetries (\ref{symmetriesVW}bc) \emph{only} for $\psi=0$ and $\pi$.
The ``point'' $\psi=0$ with $U=V$
and its ``antipode'' $\psi=\pi$ with $U=W$
will be shown to correspond to the (unique) vacuum solution 
$\mathrm{V}$ and the new sphaleron $\widehat{\mathrm{S}}$, respectively.
As mentioned above, there is, in the radial gauge,  
a three--sphere with $\widehat{\mathrm{S}}$ 
at the top ($\psi=\pi$) and $\mathrm{V}$ at the bottom  ($\psi=0$),
where the ``height'' corresponds to the energy of the configuration.

The Cartesian components of the sphaleron-like $SU(3)$ gauge field 
configuration are now defined as follows:  
\begin{equation}  
A_0(r,\theta,\phi) = 0 \,,
\quad A_m(r,\theta,\phi) = 
-g^{-1}\, f(r) \,\partial_m W(\theta,\phi)\: W^{-1}(\theta,\phi)\,,  
\label{Asphal-like}
\end{equation}
with $m=1,2,3$, and the following \bcs~on the radial function:
\begin{equation}
f(0)=0\,, \quad \lim_{r \to \infty}  f(r) =1\,.
\label{fbcs}
\end{equation}
The corresponding Higgs field configuration is given by
\begin{equation}
\Phi(r,\theta,\phi) = h(r)\, \eta\,  W(\theta,\phi) \,
\vect{1\\0\\0}\,,
\label{Phisphal-like}
\end{equation}
with the following \bcs:
\begin{equation}
h(0)=0\,, \quad \lim_{r \to \infty}  h(r) =1\,.
\label{hbcs}
\end{equation}
More specifically, the radial functions $f(r)$ and $h(r)$ are even and 
odd in $r$, with $f(r)\propto r^2$ and $h(r)\propto r$ near the origin.

For the vacuum solution $\mathrm{V}$, the matrix $W(\theta,\phi)$ in  
Eqs.~(\ref{Asphal-like}) and (\ref{Phisphal-like}) is replaced
by the constant matrix $V$ defined by Eq.~(\ref{Vdef})
and the profile function $h(r)$ in  Eq.~(\ref{Phisphal-like})
by the constant value $1$. The vacuum field configurations are then
$A_0=A_m=0$ and $\Phi=\eta\,(1,0,0)^\mathrm{\,t}$,
with $\mathrm{t}$ standing for the transpose.

The gauge and Higgs fields of Eqs.~(\ref{Asphal-like}) 
and (\ref{Phisphal-like}) are static and in the radial gauge ($A_r=0$).
They carry the basic structure of the sphaleron $\widehat{\mathrm{S}}$
but are not general enough to solve the field equations 
consistently. In the next section, we present a suitable 
generalization.

\section{Sphaleron Ansatz}
\label{sec:SphaleronAnsatz}

\subsection{Bosonic fields}
\label{sec:Bosonic fields}

As shown in the previous section,  
the matrix function $W(\theta,\phi)$ from
Eq.~(\ref{Wdef}) has an axial symmetry given by Eq.~(\ref{axialsymmVW}).
This implies that the gauge field (\ref{Asphal-like}), 
with spherical components defined by
\begin{align}  
A_\phi &\equiv -g^{-1}\, f(r)\, \partial_\phi W\, W^{-1}\,, &
A_\theta &\equiv -g^{-1}\,f(r)\, \partial_\theta W\, W^{-1}\,, & 
A_r &\equiv 0 \,,
\label{Aspherical}
\end{align}
has the property
\begin{equation} 
\partial_\phi A_c=[-2\,U_3, A_c]\,,
\label{Aaxialsymm}
\end{equation}
where $c$ stands for the component $\phi$, $\theta$, or $r$, and 
$2\,U_3$ is a constant matrix,
\begin{equation} 
-2\,U_3 \equiv -\frac{\i}{2}(\lambda_3-\sqrt{3}\,\lambda_8)
= \i \threematw{ 0 & 0 & 0 \\[-1ex] 0 & 1 & 0 \\[-1ex] 0 & 0 & -1}\,.
\end{equation}
Next, introduce the following sets of matrices:  
\begin{subequations}\label{TVUmat}    
\begin{align}
T_\phi &\equiv
-\sin\phi\,\frac{\lambda_1}{2\i}+\cos\phi\,\frac{\lambda_2}{2\i}\,,
&
T_\rho & \equiv \cos\phi\, \frac{\lambda_1}{2\i} + \sin\phi\,
\frac{\lambda_2}{2\i} \,,
\nonumber \\[2ex]
T_3 &\equiv \frac{\lambda_3}{2\i}\,, \label{Tmat}
\end{align}
\begin{align}
V_\phi &\equiv +\sin\phi\,  \frac{\lambda_4}{2\i} +\cos\phi\,
\frac{\lambda_5}{2\i}\,,
&
V_\rho &\equiv \cos\phi\, \frac{\lambda_4}{2\i}  - \sin\phi\,
\frac{\lambda_5}{2\i} \,,
\nonumber \\[2ex]
V_3 &\equiv \frac{\sqrt{3}\,\lambda_8+\lambda_3}{4\i}\,,  
 \label{Vmat}
\end{align}
\begin{align}
U_\phi &\equiv
\sin(2\phi)\,\frac{\lambda_6}{2\i}+\cos(2\phi)\,\frac{\lambda_7}{2\i}\,, 
&
U_\rho &\equiv \cos(2\phi)\,\frac{\lambda_6}{2\i}  - \sin(2\phi)\,
\frac{\lambda_7}{2\i} \,,
\nonumber \\[2ex]
U_3 &\equiv \frac{\sqrt{3}\,\lambda_8-\lambda_3}{4\i}\,, 
\label{Umat}
\end{align}
\end{subequations}
which have the same property
\begin{equation}
\partial_\phi X = [ -2 U_3, X]\,,
\end{equation}
with $X$ standing for any of the matrices defined in Eqs.~(\ref{TVUmat}abc). 
The sets $\{T_a\}$, $\{V_a\}$, $\{U_a\}$ generate the three
$su(2)$ subalgebras of $su(3)$ known as $T$--spin, $V$--spin, and $U$--spin.
Since $T_3$, $V_3$, and $U_3$ are not independent, we will use the basis
$\{T_\phi$,
$T_\rho$, $V_\phi$, $V_\rho$, $U_\phi$, $U_\rho$, $\lambda_3/(2\i)$,
$\lambda_8/(2\i)\}$ of $su(3)$ in the subsequent discussion.

The gauge field (\ref{Asphal-like}) of the approximate sphaleron
can be expanded in terms of this $su(3)$
basis. It turns out that the component $A_\phi$ involves only $T_\rho$, 
$V_\rho$, $U_\rho$, $\lambda_3/(2\i)$, $\lambda_8/(2\i)$, 
and that  $A_\theta$ involves
only $T_\phi$, $V_\phi$, $U_\phi$.
An \emph{Ansatz} which has this structure and which generalizes (\ref{Asphal-like}) 
is given by 
\mathindent=0.5em 
\begin{subequations}\label{AsphericalAnsatz} 
\begin{eqnarray} 
g\, \widehat{A}_0(r,\theta,\phi) &=& 0 \,,\\[4mm] 
g\,\widehat{A}_\phi(r,\theta,\phi) &=& 
\alpha_1(r,\theta) \,\cos\theta\; T_\rho + 
\alpha_2(r,\theta) \;V_\rho + 
\alpha_3(r,\theta) \,\cos\theta\;U_\rho + \no\\[2mm]
&& 
\alpha_4(r,\theta) \;\frac{\lambda_3}{2\i} + 
\alpha_5(r,\theta) \;\frac{\lambda_8}{2\i}  \,,\\[4mm]
g\, \widehat{A}_\theta(r,\theta,\phi) &=& 
\alpha_6(r,\theta) \;T_\phi + 
\alpha_7(r,\theta) \,\cos\theta\;V_\phi + 
\alpha_8(r,\theta) \;U_\phi \,,\\[4mm]
g\, \widehat{A}_r(r,\theta,\phi) &=&  \frac{1}{r}\: \Bigl[ \, 
\alpha_9(r,\theta)\, \cos\theta\; T_\phi + \alpha_{10}(r,\theta) \;
V_\phi +  \alpha_{11} (r,\theta)\, \cos\theta\; U_\phi \Bigr] \,, 
\end{eqnarray}
\end{subequations}
\mathindent=2em %%default for elsart.cls
with real functions $\alpha_j$, for $j=1, \ldots, 11$, depending 
on $r$ and $\theta$ and matrices $T,V,U$ from Eqs.~(\ref{TVUmat}abc)
depending implicitly on $\phi$. 
There are the following boundary conditions 
at the coordinate origin ($r=0$): 
\begin{equation}
\alpha_j(0,\theta) = 0\,,\quad \mathrm{for}\;\; j=1, \ldots, 11,
\label{alphaBCSorigin}
\end{equation}
on the symmetry axis ($\bar{\theta}=0,\pi$): 
\begin{subequations} \label{alphaBCSaxis}  
\begin{align} 
\alpha_j(r,\bar{\theta})&= 
 \bar\alpha_j(r)\, \sin\theta \,\bigr|_{\,\theta=\bar{\theta}}\,, 
 &\mathrm{for}\;\; j&=1, 2, 9, 10,
\label{alphaBCSaxis12910}\\[2mm]
\alpha_j(r,\bar{\theta})&= 
 \bar\alpha_j(r)\, \sin^2\theta \,\bigr|_{\,\theta=\bar{\theta}}\,, 
&\mathrm{for}\;\; j&=3,4,5,11,
\label{alphaBCSaxis34511}\\[2mm]
\alpha_j(r,\bar{\theta})&=
 (-)^{j-5}\,\cos\theta\,\partial_\theta\,
\alpha_{j-5}(r,\theta)\,\bigr|_{\,\theta=\bar{\theta}}\,, 
&\mathrm{for}\;\; j&=6,7,
\label{alphaBCSaxis67}\\[2mm]
\alpha_j(r,\bar{\theta})&=
 \half\,\cos\theta\,\partial_\theta\,
\alpha_{j-5}(r,\theta)\,\bigr|_{\,\theta=\bar{\theta}}\,, 
&\mathrm{for}\;\; j&=8,
\label{alphaBCSaxis678} 
\end{align}
\end{subequations}  
and towards infinity:
\begin{equation} 
\lim_{r \to \infty}  \,
\left(
\begin{array}{c}
\alpha_1(r,\theta)\\
\alpha_2(r,\theta)\\
\alpha_3(r,\theta)\\
\alpha_4(r,\theta)\\
\alpha_5(r,\theta)\\
\alpha_6(r,\theta)\\
\alpha_7(r,\theta)\\
\alpha_8(r,\theta)\\
\alpha_9(r,\theta)\\
\alpha_{10}(r,\theta)\\
\alpha_{11}(r,\theta)
\end{array}
\right)
= 
\left(
\begin{array}{c}
-2\,\sin\theta\,(1+\sin^2\theta)  
\\   
2\,\sin\theta\,\cos^2\theta\\
- 2\sin^2\theta\\
-\sin^2\theta\,(1+2\,\sin^2\theta)  
\\ 
\sqrt{3}\,\sin^2\theta\\
2\\
2\\
-2\,\sin\theta\\
0\\
0\\
0
\end{array}
\right)\,.
\label{alphaBCSinfinity}
\end{equation}
Furthermore, the $\alpha_j$ are required to have positive parity 
with respect to reflection of the $z$-coordinate,
\begin{equation}
\alpha_j(r,\pi-\theta) = +\alpha_j(r,\theta)\,,
\quad \mathrm{for}\;\; j=1, \ldots, 11,
\label{alphajparity}
\end{equation}
in order to keep the reflection symmetry of the approximate sphaleron 
configuration; cf. Eqs.~(\ref{reflsymW}) and (\ref{Asphal-like}). 
For completeness, the Cartesian components of the gauge field \emph{Ansatz} are
\begin{subequations}   
\begin{align}
g\, \widehat{A}_1 =&   -\,\frac{\sin\phi}{\rho} 
\Bigl[ \,\alpha_1 \cos\theta\, T_\rho + \alpha_2
  V_\rho + \alpha_3\cos\theta\, U_\rho + \alpha_4 \lambda_3/(2\i) +
  \alpha_5 \lambda_8/(2\i) \,\Bigr] \no\\
& + \,\frac{\sin\theta\,\cos\phi}{\rho} \Bigl[ \,\cos\theta
  (\alpha_6+\sin\theta\,\alpha_9)\, T_\phi + (\cos^2\theta\, \alpha_7
  +\sin\theta\, \alpha_{10})\, V_\phi \no \\
&  +\,\cos\theta (\alpha_8+\sin\theta\,
  \alpha_{11}) \, U_\phi \,\Bigr]\,,
\\[2mm]
g\, \widehat{A}_2 =& 
\frac{\cos\phi}{\rho} \Bigl[ \,\alpha_1 \cos\theta\, T_\rho + \alpha_2
  V_\rho + \alpha_3\cos\theta\, U_\rho + \alpha_4 \lambda_3/(2\i) +
  \alpha_5 \lambda_8/(2\i) \,\Bigr] \no\\
&  + \,\frac{\sin\theta\,\sin\phi}{\rho} \Bigl[ \,\cos\theta\, (\alpha_6 + \sin\theta \,
  \alpha_9)\, T_\phi + (\cos^2\theta  \,\alpha_7 +\sin\theta \,
  \alpha_{10})\, V_\phi \no \\
&  +\,\cos\theta (\alpha_8 +\sin\theta  \, \alpha_{11})
  \, U_\phi \,\Bigr]\,,
\\[2mm]
g\, \widehat{A}_3 =& 
-\, \frac{\cos\theta}{z} \Bigl[ \,(\sin\theta\,\alpha_6 
-\cos^2\theta\, \alpha_9)\,
  T_\phi + \cos\theta\, (\sin\theta\, \alpha_7 -\alpha_{10})\, V_\phi 
\no \\
&  + \,(\sin\theta \,\alpha_8 -\cos^2\theta \, \alpha_{11})\, U_\phi
\,\Bigr] \,,
\end{align}
\end{subequations}
with a $T$--spin structure similar to the $\mathrm{S}$ and  
$\mathrm{S}^\star$ \emph{Ans\"{a}tze} \cite{K93}.

Turning to the Higgs field (\ref{Phisphal-like}) of the approximate 
sphaleron, we observe that
the direction of the vacuum expectation value has been chosen so that 
$2\,U_3 \Phi_{\rm vac}=0$. 
The field $\Phi$ is found to have an axial symmetry,
\begin{equation}
\partial_\phi \Phi = -2\, U_3\, \Phi\,,
\label{Phiaxialsymm}
\end{equation}
which matches the axial symmetry (\ref{Aaxialsymm}) of the gauge 
field. 
The structure of this specific Higgs field is now generalized to the
following \emph{Ansatz}:
\begin{eqnarray} 
\widehat{\Phi}(r,\theta,\phi) 
&=& \eta\, \bigl[\, \beta_1(r,\theta)\,\lambda_3 
+\beta_2(r,\theta)\,\cos\theta\; \,2\i\,T_\rho
+\beta_3(r,\theta)\,2\i\,V_\rho \,\bigr]\, \vect{1\\0\\0} \no\\
&=& \eta\,\vectleft{
\beta_1(r,\theta)\\ 
\beta_2(r,\theta)\,\cos\theta\; \e^{\i\phi} \\ 
\beta_3(r,\theta)\;\e^{-\i\phi}}\,,
\label{PhiAnsatz}
\end{eqnarray}
with real functions $\beta_k$, for $k=1,2,3$, 
depending on $r$ and $\theta$
and matrices $T_\rho$ and $V_\rho$ depending implicitly on $\phi$. 
The \bcs~are at the coordinate origin ($r=0$): 
\begin{equation}
\partial_\theta\, \beta_1(0,\theta) = 
\beta_2(0,\theta) = 
\beta_3(0,\theta) = 0\,,
\label{betaBCSorigin} 
\end{equation}
on the symmetry axis ($\bar{\theta}=0,\pi$):  
\begin{equation} 
\partial_\theta\,\beta_1(r,\theta)\,\bigr|_{\,\theta=\bar{\theta}}=0\,,\quad
\beta_k(r,\bar{\theta})= 
\bar\beta_k(r) \,\sin\theta\,\bigr|_{\,\theta=\bar{\theta}}\,,
\quad \mathrm{for}\;\; k=2,3,
\label{betaBCSaxis}
\end{equation}
and towards infinity:
\begin{align} 
\lim_{r \to \infty}  \,
\left(
\begin{array}{c}
\beta_1(r,\theta)\\
\beta_2(r,\theta)\\
\beta_3(r,\theta)
\end{array}
\right)
= 
\left(
\begin{array}{c}
\cos^2\theta\\
-\sin\theta\\
-\sin\theta
\end{array}
\right)\,.
\label{betaBCSinfinity}
\end{align}
Furthermore, these functions must be even under reflection of the 
$z$-coordinate,
\begin{align}
\beta_k(r,\pi-\theta) &= +\beta_k(r,\theta)\,,
\quad \mathrm{for}\;\; k=1,2,3.
\label{betakparity}
\end{align}

This completes the construction of the \emph{Ansatz} for the 
bosonic fields. Before discussing the resulting 
energy and field equations, the following three remarks may be helpful. 
First, the gauge  field (\ref{Asphal-like}) and Higgs field  
(\ref{Phisphal-like}) of the approximate sphaleron are 
reproduced by the following \emph{Ansatz} functions:
\begin{subequations}
\begin{eqnarray} 
\left(
\begin{array}{c}
\alpha_1(r,\theta)\\
\alpha_2(r,\theta)\\
\alpha_3(r,\theta)\\
\alpha_4(r,\theta)\\
\alpha_5(r,\theta)\\
\alpha_6(r,\theta)\\
\alpha_7(r,\theta)\\
\alpha_8(r,\theta)\\
\alpha_9(r,\theta)\\
\alpha_{10}(r,\theta)\\
\alpha_{11}(r,\theta)
\end{array}
\right)
&=& 
\,f(r)\,
\left(
\begin{array}{c}
-2\,\sin\theta\,(1+\sin^2\theta)  
\\   
2\,\sin\theta\,\cos^2\theta\\
- 2\sin^2\theta\\
-\sin^2\theta\,(1+2\,\sin^2\theta)  
\\ 
\sqrt{3}\,\sin^2\theta\\
2\\
2\\
-2\,\sin\theta\\
0\\
0\\
0
\end{array}
\right)\,,
\label{alphasfromW}
\end{eqnarray}
\begin{eqnarray} 
\left(\begin{array}{c}  
\beta_1(r,\theta)\\
\beta_2(r,\theta)\\
\beta_3(r,\theta)
\end{array}\right)
&=& 
\,h(r)\, 
\left(
\begin{array}{c}
\cos^2\theta\\
-\sin\theta\\
-\sin\theta
\end{array}\right)\,,
\label{betasfromW}
\end{eqnarray}
\end{subequations}
which explains in part the choice 
of \bcs~(\ref{alphaBCSorigin})--(\ref{alphaBCSinfinity}) and 
(\ref{betaBCSorigin})--(\ref{betaBCSinfinity})
for the general \emph{Ansatz}.

Second, the \emph{Ansatz} as given by Eqs.~(\ref{AsphericalAnsatz})
and (\ref{PhiAnsatz}) has a residual $SO(3)$ gauge symmetry under the 
following transformations:
\begin{subequations}
\begin{equation}  
g\, \widehat{A}^\prime_n = 
\Omega\, \left(\,g\,\widehat A_n + \partial_n \,\right)\,\Omega^{-1}\,, \quad
\widehat \Phi^\prime = \Omega\, \widehat\Phi\,,
\label{ResidualSymmetry}
\end{equation}
with
\begin{equation}
\Omega(r,\theta,\phi) \equiv 
\exp\bigl[\,\omega_T(r,\theta)\, T_\phi + \omega_V(r,\theta)\,
V_\phi + \omega_U(r,\theta)\, U_\phi\,\bigr]\,,
\end{equation}
\end{subequations}
for two-dimensional  parameter functions $\omega_s(r,\theta)$, $s=T,V,U$.

Third, the structure of the \emph{Ansatz} gauge 
field (\ref{AsphericalAnsatz}) is quite intricate:
for a given halfplane through the $z$--axis with azimuthal 
angle $\phi$, the parallel components
$\widehat{A}_r$ and $\widehat{A}_\theta$ 
involve only one particular $su(2)$ subalgebra of $su(3)$,
whereas the orthogonal  component $\widehat{A}_\phi$
excites precisely the other five generators of $su(3)$.

\subsection{Energy and field equations}
\label{sec:Energy}

The bosonic energy functional of theory (\ref{actionYMHW}) is given by
\begin{equation}
E[A,\Phi] = \int_{\bR^3} \dvx \left[\, -\frachalf \, {\rm tr} (F_{mn})^2 +
  |D_m\Phi|^2 +\lambda \left( |\Phi|^2-\eta^2\right)^2 \,\right] 
  \,,
\label{energyfunctional}
\end{equation}
for spatial indices $m,n=1,2,3$.
From the sphaleron \emph{Ans\"{a}tze} (\ref{AsphericalAnsatz}) and
(\ref{PhiAnsatz}), one obtains
\begin{equation}
E[\widehat{A},\widehat{\Phi}] = 
4\pi \int_{0}^{\infty} \dint{r} \int_{0}^{\pi/2}\dint{\theta}
\, r^2\sin\theta\;\, \widehat{e}(r,\theta) \,,
\label{Eansatz}
\end{equation}
where the energy density $\widehat{e}(r,\theta)$ contains contributions from the
Yang--Mills term, the kinetic Higgs term, and the Higgs potential in the
energy functional, 
\begin{align}
\widehat{e}(r,\theta) &= 
\widehat{e}_{\rm \,YM}(r,\theta) + \widehat{e}_{\rm \,Hkin}(r,\theta) 
+ \widehat{e}_{\rm  \,Hpot}(r,\theta)\,.
\label{edens}
\end{align}
The detailed expressions for these energy density contributions 
are relegated to Appendix~\ref{sec:AppendixA}.
The energy density (\ref{edens})
turns out to be well-behaved due to the 
\bcs~(\ref{alphaBCSorigin})--(\ref{alphaBCSinfinity})
and (\ref{betaBCSorigin})--(\ref{betaBCSinfinity}).
In addition, there is a reflection symmetry,     
$\widehat{e}(r,\theta)=\widehat{e}(r,\pi-\theta)$,
which allows the range of $\theta$ in Eq.~(\ref{Eansatz}) to be 
restricted to $[0,\pi/2]$.

It can be verified that the \emph{Ans\"{a}tze}  
(\ref{AsphericalAnsatz})--(\ref{alphajparity}) and 
(\ref{PhiAnsatz})--(\ref{betakparity})
consistently solve the field equations of the theory
(\ref{sec:Theory}). That is, they reproduce the variational
equations from the \emph{Ansatz} energy functional
(\ref{Eansatz}). This result is a manifestation of the
principle of symmetric criticality \cite{P79}, which 
states that in the quest of stationary points it suffices,
under certain conditions, to consider variations that respect
the symmetries of the \emph{Ansatz} (here, rotation and reflection symmetries).

It is difficult to prove that the resulting system of partial differential
equations for the \emph{Ansatz} functions $\alpha_j(r,\theta)$ and 
$\beta_k(r,\theta)$, together with the conditions
(\ref{alphaBCSorigin})--(\ref{alphajparity}) and 
(\ref{betaBCSorigin})--(\ref{betakparity}), has a nontrivial solution, 
that is, a non-vacuum solution.
It is, however, possible to give a heuristic argument in favor of a
nontrivial \emph{regular} solution 
(a similar argument has been used for $\mathrm{S}^\star$ \cite{K93}).

According to the boundary condition (\ref{betaBCSinfinity}),
the function $\beta_1(r,\theta)$ vanishes asymptotically for $\theta=\pi/2$, 
i.e., on the equatorial circle at infinity. 
Let us simply exclude the case of an ``isolated''
zero of $\beta_1(r,\theta)$  at $(r,\theta)=(\infty,\pi/2)$,
which is discussed further in Appendix~\ref{sec:AppendixB}.   
For a regular solution and by reflection symmetry $\theta \to \pi - \theta$, 
the set of zeroes of $\beta_1(r,\theta)$ then
extends inwards from infinity along the equatorial plane 
$\theta=\pi/2$ and intersects the symmetry axis of the \emph{Ansatz} 
at the coordinate origin $r=0$.
Since $\beta_2(r,\theta)$ and $\beta_3(r,\theta)$  also vanish due to
the boundary conditions (\ref{betaBCSaxis}), the Higgs field 
(\ref{PhiAnsatz}) is exactly zero at this point, $\widehat\Phi(0,0,0)=0$, 
which is not possible for vacuum configurations with 
$|\Phi(r,\theta,\phi)|=\eta$.

It appears that the details of the (regular) solution
can  only be determined 
by a numerical evaluation of the reduced field equations.
The explicit numerical solution of these partial differential equations
is, however, rather difficult.
In the following, we will simply use the
approximate sphaleron fields (\ref{Asphal-like}) and (\ref{Phisphal-like}), 
which are sufficient for our purpose of looking for
fermion zero modes.\footnote{The simplified fields
(\ref{Asphal-like}) and (\ref{Phisphal-like}), 
with appropriate radial functions
$f(r)$ and $h(r)$, can also be used to get an upper 
bound on the energy of $\widehat{\mathrm{S}}$.
For the case of $\lambda/g^2=0$, one obtains  
$E[\,\widehat{\mathrm{S}}\,]  < 1.72\times E[\,\mathrm{S}\,]$,
where $E[\,\mathrm{S}\,]$ denotes the energy  of the $SU(2)$ 
sphaleron $\mathrm{S}$ \cite{KM84} embedded in the $SU(3)$ \YMHth.
This implies that 
$E[\,\widehat{\mathrm{S}}\,] < E[\,\mathrm{S}^\star\,]$,
at least for $\lambda/g^2=0$, where $E[\,\mathrm{S}^\star\,] \approx
1.91\times E[\,\mathrm{S}\,]$ 
denotes the energy of the embedded $SU(2)$ 
sphaleron $\mathrm{S}^\star$ \cite{K93}.}

\section{Fermion zero modes}
\label{sec:Fermionzeromodes}

\subsection{Ansatz and zeromode equation}
\label{sec:Ansatzandzeromodeequation}

As mentioned in Section~\ref{sec:Theory}, the $SU(3)$ gauge theory 
considered has a single massless left-handed  fermionic field 
in the $\mathbf{3}$ representation.  
This fermionic field  is now taken to interact with
the fixed background gauge field 
of the approximate sphaleron (\ref{Asphal-like}) 
or the general sphaleron \emph{Ansatz} (\ref{AsphericalAnsatz}). 
In both cases,   
the corresponding fermion Hamiltonian, 
\begin{equation}
H=\i \vec \sigma \cdot (\nabla-g \vec A) \,,
\end{equation}
has the following symmetries:
\begin{equation}
[K_3,H]=0\,,\quad \{R_3,H\}=0\,,
\end{equation}
with
\begin{subequations}
\begin{align}
K_3 &\equiv -\i \pr_\phi + \frachalf \left(  \lambda_3-\sqrt{3}\,\lambda_8\right)
+ \frachalf \sigma_3\,, \label{K3def} \\[2mm]
R_3 &\equiv 
{\rm diag}(-1,+1,-1,+1,-1,+1)
\, \mathcal{R}_3\,, \label{R3def}
\end{align}
\end{subequations}
and $\mathcal{R}_3$ the coordinate reflection operator with respect to  the
3-axis, i.e., $\mathcal{R}_3\, \theta = \pi-\theta$, $\mathcal{R}_3\,
\phi=\phi$, and $\mathcal{R}_3\, r=r$.
The $\lambda$ and $\sigma$ matrices in  
Eq.~(\ref{K3def}) operate, of course, on different 
indices of the fermionic field, 
the $SU(3)$  ``color'' and $SU(2)$  spin indices 
respectively (see below).  
Observe that $R_3$ and $K_3$ commute, $[R_3,K_3]=0$.

The eigenvalue $\kappa$ of $K_3$
can take odd-half-integer values ($\kappa \in \bZ+\frachalf$) and 
the corresponding eigenstate has the form
\begin{equation}
\widehat \Psi(r,\theta,\phi) = \vect{
F_{R+}(r,\theta)\,\, \e^{\i\, (\kappa-1/2)\, \phi}\\[1ex]
F_{G+}(r,\theta)\,\, \e^{\i\, (\kappa+1/2)\, \phi}\\[1ex]
F_{B+}(r,\theta)\,\, \e^{\i\, (\kappa-3/2)\, \phi}\\[1ex]
F_{R-}(r,\theta)\,\, \e^{\i\, (\kappa+1/2)\, \phi}\\[1ex]
F_{G-}(r,\theta)\,\, \e^{\i\, (\kappa+3/2)\, \phi}\\[1ex]
F_{B-}(r,\theta)\,\, \e^{\i\, (\kappa-1/2)\, \phi}
} \,,
\label{K3eigenstate}
\end{equation}
where $+$ and $-$ denote the spin (in units of $\hbar/2$)
and $R$, $G$, and $B$ the color (red, green, and blue).
The real functions  $F_{cs}(r,\theta)$ 
must vanish for $r\to\infty$,
in order to have a normalizable solution,
\begin{equation} 
\int_{0}^{\infty} \dint{r} \int_{0}^{\pi}\dint{\theta}
\int_{0}^{2\pi}\dint{\phi}
\, r^2\sin\theta\;\,|\widehat{\Psi}|^2 = 1 \,.
\label{ZMnorm}
\end{equation}
Those functions $F_{cs}(r,\theta)$ 
which are accompanied by a 
$\phi$-dependent phase factor in Eq.~(\ref{K3eigenstate})
are required to vanish also for $\theta=0, \pi$ and $r=0$, 
in order to have regular field configurations.

For the gauge field (\ref{Asphal-like}) of the approximate sphaleron and a 
particular eigenvalue $\kappa$ of the \emph{Ansatz} (\ref{K3eigenstate}), 
the zeromode equation $H\, \widehat{\Psi}=0$ reduces to the following 
partial differential equation:
\begin{equation}
\left\{ A_1(\theta)\, \pdq{}{r} + \frac{1}{r} \, A_2(\theta)\,
  \pdq{}{\theta} + \frac{1}{r}\, A_3(\kappa)\, + \frac{f(r)}{r} \, A_4(\theta)
  \right\} \, \widehat \Psi(r,\theta,0) =0\,,
\label{zeromodeeqn}
\end{equation}
with matrices
\mathindent=0em 
\begin{subequations}    
\begin{align}
A_1(\theta) &= \twomat{ \sin\theta\,\cos\theta\, \id_3 
&\;\; \sin^2\theta   \, \id_3\\[1ex]
\sin^2\theta\, \id_3 
&\;\; -\sin\theta\, \cos\theta\,
  \id_3}  \label{A1def} \,,
\end{align}
\begin{align}
A_2(\theta) &=  \twomat{ -\sin^2\theta \, \id_3 
&\;\;  \sin\theta\,\cos\theta \, \id_3   \label{A2def} \,, \\[1ex]
\sin\theta\,\cos\theta \, \id_3 
&\;\; \sin^2\theta \, \id_3 }\,, 
\end{align}
\begin{align}
A_3(\kappa) &= \sixmat{ 0&0&0 & \kappa+1/2 & 0 & 0  \\
0&0&0& 0& \kappa+3/2 & 0\\
0&0&0& 0&0& \kappa -1/2\\
-\kappa + 1/2 & 0&0 &0&0&0\\
0 & -\kappa -1/2 & 0 & 0&0&0\\
0&0&-\kappa +3/2 &0&0&0}   \label{A3def} \,, 
\end{align}
\begin{align}
A_4(\theta) &= \twomat{
   A_4^{(1)}(\theta) 
   &\;\; A_4^{(2)}(\theta) \\[1ex]
   -A_4^{(2)\,\mathrm{t}}(\theta) 
   &\;\; A_4^{(1)\,\mathrm{t}}(\theta)}   \label{A4def} \,, 
\end{align}
\begin{align}
A_4^{(1)}(\theta) &= \threemat{
0 
&\;\; \sin^2\theta
&\;\; \cos\theta \sin^2\theta \\
-\sin^2\theta 
&\;\; 0 
&\;\; -\sin^3\theta \\
-\sin^2\theta \cos\theta  
&\;\;  \sin^3\theta 
&\;\; 0 } \,, 
\end{align}
\begin{align}
A_4^{(2)}(\theta) &= \threemat{
\sin^4\theta 
&\;\; \sin^3\theta \cos\theta 
&\; -2\sin\theta \cos^2\theta  \\[1ex]
\sin\theta \cos\theta\, \left( 3-\cos^2\theta\right)
&\;\; \sin^2\theta\, \left(\cos^2\theta-2\right) 
&\; 2\cos\theta \sin^2\theta\\[1ex]
0 
&\;\; 0 
&\; \sin^2\theta},
\end{align}
\end{subequations}
\mathindent=2em 
where the superscript $\mathrm{t}$ in (\ref{A4def}) indicates the transpose.
Fermion zero modes correspond to \emph{normalizable} solutions
of Eq.~(\ref{zeromodeeqn}).  

Since $R_3$ anticommutes with $H$,  $\{R_3,H\}=0$,
common eigenstates with zero energy can be found. 
As shown in Appendix~\ref{sec:AppendixC}, zero-energy eigenstates 
with opposite eigenvalues of $K_3$ and $R_3$
can be constructed if one starts from
a generator $\widetilde U$ of 
$\pi_5[SU(3)]$ 
with opposite winding number compared to $U$.

\subsection{Asymptotic behavior}
\label{sec:Asymptoticbehavior}

The gauge field function $f(r)$ of the 
sphaleron-like field (\ref{Asphal-like}) is approximately equal to one
for large enough $r$. 
With $f=1$ exactly, the Yang--Mills field is pure gauge, 
\begin{equation} 
A_m= -g^{-1}\,\pr_m W\,  W^{-1}\,,
\end{equation} 
in terms of the $SU(3)$ matrix $W(\theta,\phi)$ defined by Eq.~(\ref{Wdef}). 
The Dirac Hamiltonian becomes then
\begin{equation} 
H = \i \vec \sigma \cdot 
\bigl( \nabla -g \vec A \,\bigr) = W\, H_{\rm free}\, W^{-1}\,,  
\end{equation}
with $H_{\rm free} \equiv \i \vec \sigma \cdot \nabla$. 
Asymptotically, the normalizable
zero mode can therefore be written as
\begin{equation} 
\widehat \Psi^{(\infty)}(r,\theta,\phi) = 
W(\theta,\phi)\, \Psi_{\rm free}(r,\theta,\phi)\,, 
\end{equation}
for an appropriate solution of $H_{\rm free}\, \Psi_{\rm free}=0$.

The wave function
$\widehat \Psi^{(\infty)}$ is an eigenstate of $K_3$, 
if $\Psi_{\rm free}$ is an eigenstate of 
$\widetilde K_3$ with the same eigenvalue, where $\widetilde K_3$ 
is defined by 
$\widetilde K_3 \equiv -\i \partial_\phi 
 - \frachalf (\lambda_3-\sqrt{3}\,\lambda_8)+ \frachalf \sigma_3$.
Since $H_{\rm free}$ does not mix colors, the asymptotic two-spinors can
be given separately for each color,
\mathindent=0.5em 
\begin{equation} 
\psi_{\rm free}^{(c)}(r,\theta,\phi) = \frac{1}{r^{l+1}} 
\vect{
+\sqrt{(l+1/2-\kappa-c)/(2l+1)}
\,\; Y_l^{\kappa+c-1/2}(\theta,\phi) \\[2mm]
-\sqrt{(l+1/2+\kappa+c)/(2l+1)} 
\,\; Y_l^{\kappa+c+1/2}(\theta,\phi)}.
\end{equation}
\mathindent=2em 
Here, $\kappa$ is the $\widetilde K_3$--eigenvalue, 
$l$ stands for the orbital angular-momentum quantum number, 
and $c$ takes the values $0,-1,+1$ 
for the colors $R,G,B$, respectively.
Most importantly, the quantum number $l$ is restricted by the condition
\begin{equation}
 -l \le \kappa+c-\frachalf \le l-1\,,
\end{equation}
which ensures the consistency of the orbital angular momentum 
and the $\phi$-dependence from the $K_3$-symmetry.

The asymptotic $K_3$--eigenstate $\widehat \Psi^{(\infty)}(r,\theta,\phi)$ 
with eigenvalue $\kappa$ is then given by 
\mathindent=0em 
\begin{equation}
\widehat \Psi^{(\infty)} = W \cdot \vectleft{
\phantom{-}
\sum_{l=|\kappa|+\frachalf\phantom{-1}}^\infty D_{R}^{(l)}\,r^{-(l+1)} \,\sqrt{
  (l+\frachalf-\kappa)/(2l+1)} \; Y_l^{\kappa-1/2} \\[3ex]
\phantom{-}
\sum_{l=|\kappa-1|+\frachalf}^\infty D_{G}^{(l)}\,r^{-(l+1)} \,\sqrt{
  (l+\tfr{3}{2}-\kappa)/(2l+1)} \; Y_l^{\kappa-3/2} \\[3ex]
\phantom{-}
\sum_{l=|\kappa+1|+\frachalf}^\infty D_{B}^{(l)}\,r^{-(l+1)} \,\sqrt{
  (l-\tfr{1}{2}-\kappa)/(2l+1)} \; Y_l^{\kappa+1/2} \\[3ex]
-\sum_{l=|\kappa|+\frachalf\phantom{-1}}^\infty D_{R}^{(l)}\,r^{-(l+1)} \,\sqrt{
  (l+\frachalf+\kappa)/(2l+1)} \; Y_l^{\kappa+1/2} \\[3ex]
-\sum_{l=|\kappa-1|+\frachalf}^\infty D_{G}^{(l)}\,r^{-(l+1)} \,\sqrt{
  (l-\tfr{1}{2}+\kappa)/(2l+1)} \; Y_l^{\kappa-1/2} \\[3ex]
-\sum_{l=|\kappa+1|+\frachalf}^\infty D_{B}^{(l)}\,r^{-(l+1)} \,\sqrt{
  (l+\tfr{3}{2}+\kappa)/(2l+1)} \; Y_l^{\kappa+3/2}
},
\label{Psihat-infinity}
\end{equation}
with constant coefficients $D_R^{(l)}$, $D_G^{(l)}$, and $D_B^{(l)}$.
This implies an asymptotic $r^{-2}$ 
behavior for the fermion zero modes, provided the sums 
in (\ref{Psihat-infinity}) converge and are nonvanishing.

For $r$ close to zero, one has $f(r) \propto r^2$
from the requirement of finite Yang--Mills energy density. 
With $f=0$ exactly, the gauge field vanishes, $A_m=0$.  
The  behavior at the origin is then given by
\begin{equation}
\widehat \Psi^{(0)} = \vectleft{
\sum_{l=|\kappa|+\frachalf\phantom{-1}}^\infty C_R^{(l)}\, r^l \,\sqrt{
 (l+\tfr{1}{2}+\kappa)/(2l+1)}\; Y_l^{\kappa-1/2}  
+ C_R^{(0)} \, \delta_{\kappa-\frac{1}{2},0}\\[3ex]
\sum_{l=|\kappa+1|+\frachalf}^\infty C_G^{(l)}\, r^l \,\sqrt{
 (l+\tfr{3}{2}+\kappa)/(2l+1)} \; Y_l^{\kappa+1/2} 
+ C_G^{(0)} \, \delta_{\kappa+\frac{1}{2},0}\\[3ex]
\sum_{l=|\kappa-1|+\frachalf}^\infty C_B^{(l)}\, r^l \,\sqrt{
 (l-\frachalf+\kappa)/(2l+1)}\; Y_l^{\kappa-3/2}
+ C_B^{(0)} \, \delta_{\kappa-\frac{3}{2},0}\\[3ex]
\sum_{l=|\kappa|+\frachalf\phantom{-1}}^\infty C_R^{(l)}\, r^l \,\sqrt{
 (l+\tfr{1}{2}-\kappa)/(2l+1)}\; Y_l^{\kappa+1/2}  
+ C_R^{(0)} \, \delta_{\kappa+\frac{1}{2},0}\\[3ex]
\sum_{l=|\kappa+1|+\frachalf}^\infty C_G^{(l)}\, r^l \,\sqrt{
 (l-\tfr{1}{2}-\kappa)/(2l+1)} \; Y_l^{\kappa+3/2} 
+ C_G^{(0)} \, \delta_{\kappa+\frac{3}{2},0}\\[3ex]
\sum_{l=|\kappa-1|+\frachalf}^\infty C_B^{(l)}\, r^l \,\sqrt{
 (l+\tfr{3}{2}-\kappa)/(2l+1)}\; Y_l^{\kappa-1/2}
+ C_B^{(0)} \, \delta_{\kappa-\frac{1}{2},0}\\[3ex]
},
\label{Psihat-origin}
\end{equation}
\mathindent=2em 
with constant coefficients $C_R^{(l)}$, $C_G^{(l)}$, and $C_B^{(l)}$.
The wave function $\widehat \Psi^{(0)}(r,\theta,\phi)$ 
is a $K_3$--eigenstate with eigenvalue $\kappa$.

An explicit solution of the zeromode equation (\ref{zeromodeeqn})
would relate the coefficients $D$ and $C$ of 
Eqs.~(\ref{Psihat-infinity}) and (\ref{Psihat-origin}). An analytic 
solution is, however, not feasible, possibly except on the symmetry 
axis.

\subsection{Solutions on the symmetry axis}
\label{sec:symmetryaxis}

The zeromode equations can be solved on the symmetry axis 
(coordinate $z\equiv r\,\cos\theta$) for the background  gauge 
field (\ref{Asphal-like}) of the approximate sphaleron.
Two zero modes have been found, one with $\kappa=+1/2$ and one 
with $\kappa=-1/2$.

\par
For $\kappa=-1/2$, we can, in fact, construct an analytic solution. 
On the $z$--axis, 
only the $\phi$-independent components of
the general \emph{Ansatz} (\ref{K3eigenstate}) are nonzero. 
This leads to the simplified \emph{Ansatz}
\begin{equation}
\widehat\Psi^{(-)}(0,0,z) = 
\frac{1}{\sqrt{2}}\;
\vect{0 \\ l_2^{(-)}(z)\\ 0 \\ l_4^{(-)}(z) \\ 0 \\ 
0}\,,
\label{zmminushalf}
\end{equation}
for cylindrical coordinates $(\rho,\phi,z)$
and with even functions $l_2^{(-)}(z)$ and $l_4^{(-)}(z)$. 
The wave function $\widehat \Psi^{(-)}$ is an $R_3$--eigenstate with 
eigenvalue $1$. Inserting (\ref{zmminushalf}) into the zeromode equation
(\ref{zeromodeeqn}) with all $\theta$ derivatives set to zero, 
one obtains the normalizable solution
\begin{equation}
l_2^{(-)}(z)=l_4^{(-)}(z)= l(z)\,,
\end{equation}
in terms of the following function of $z$:
\begin{equation}
l(z) \equiv\exp \left( -\int_0^{|z|} {\rm d}z'\, \frac{2f(z')}{z'} \right)\,.
\end{equation}
Hence,  $|\widehat \Psi^{(-)}|$
is normalized to one at the origin and drops off like $|z|^{-2}$ for
large $|z|$. Note that the same exponential factor appears in the
fermion zero mode of the spherically symmetric
$SU(2)$ sphaleron S \cite{N75,KM84aps,BK85,R88,KB93}.
Similar analytic solutions of fermion zero modes
on the symmetry axis have also been found for
the axially symmetric $SU(2)$ 
constrained instanton $\mathrm{I}^\star$ \cite{KW96,K98zm}.

For $\kappa=+1/2$ and $\theta$ close to $0$ or $\pi$, 
we use the \emph{Ansatz}
\begin{equation}
\widehat\Psi^{(+)}(r,\theta,\phi) = 
\vectleft{ l_1^{(+)}(r)\: \cos\theta \\
l_2^{(+)}(r)\: \sin\theta\: \e^{\i \phi} \\
l_3^{(+)}(r)\: \sin\theta\:  \cos\theta\:\e^{-\i\phi}\\
l_4^{(+)}(r)\: \sin\theta\: \e^{\i\phi}\\
l_5^{(+)}(r)\: \sin^2\theta\, \cos\theta\: \e^{2\i\phi}\\
l_6^{(+)}(r)}\,.
\label{zmplushalf}
\end{equation}
The wave function
$\widehat \Psi^{(+)}$ is an $R_3$--eigenstate with eigenvalue $1$
which 
is normalized to one at the origin, provided $l_6^{(+)}(0)=1$ and
$l_n^{(+)}(0)=0$ for $n=1, \ldots ,5$.
The differential equation (\ref{zeromodeeqn}) can then be solved numerically 
to first and second order in $\theta$ or $\pi-\theta$. 
On the symmetry axis, only the first and last entries of (\ref{zmplushalf})
are nonzero, but the functions $l_2^{(+)}, \ldots, l_5^{(+)}$
still affect the equations for $l_1^{(+)}$ and $l_6^{(+)}$ because 
of the coupling through the $\theta$ derivatives in (\ref{zeromodeeqn}).
On the $z$--axis, one has asymptotically 
$|\widehat \Psi^{(+)}| \propto |z|^{-2}$
for $(l_1,l_2,l_3,l_4,l_5,l_6)\propto (-1,1,0,0,0,1)/r^2$.

\begin{figure}[t]
\begin{center}
\includegraphics[width=10cm]{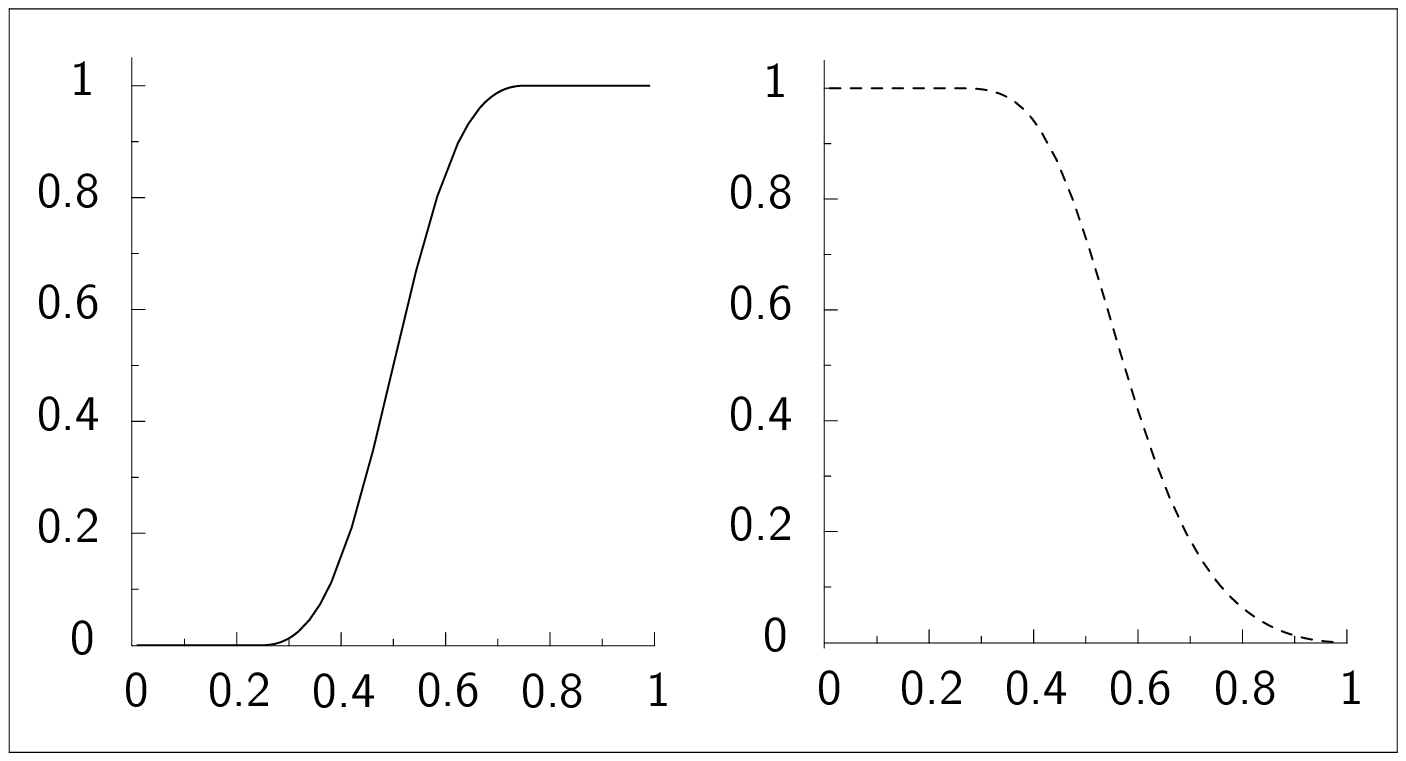}
\end{center}
\caption{Left panel: radial function $f(\zeta)$ from Eq.~(\ref{fnum})
for the $SU(3)$ background gauge field (\ref{Asphal-like}),
with $f(\zeta)$ shown 
as a function of the compact coordinate $\zeta \equiv r/(r+1)$.
Right panel: functions of the corresponding $\kappa=-1/2$ fermion zero mode
(\ref{zmminushalf}) on the symmetry axis ($r=|z|$), with 
$l_2^{(-)}(\zeta)=l_4^{(-)}(\zeta)$ shown as the broken curve.}
\label{FIG-ZMminus}
\end{figure}
\begin{figure}
\begin{center}
\includegraphics[width=10cm]{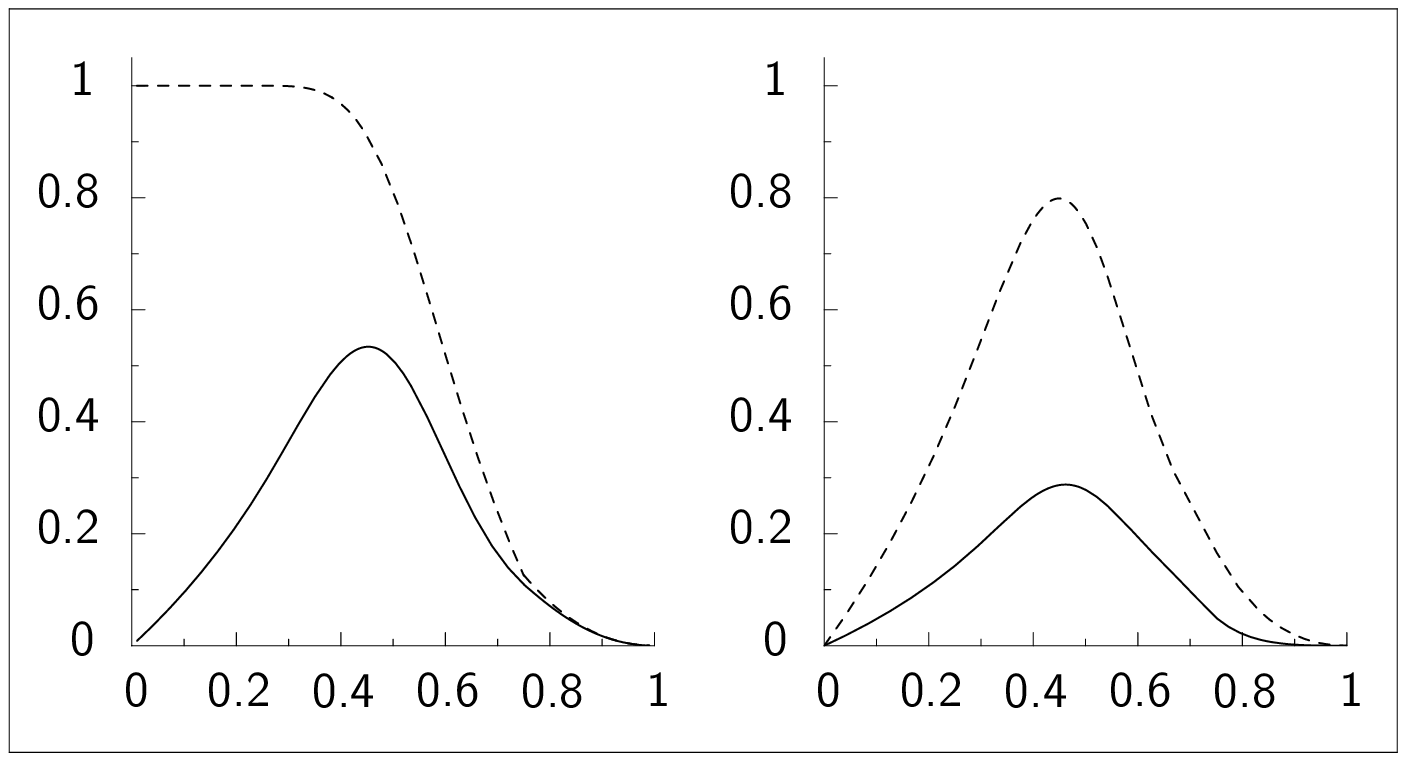}
\end{center}
\caption{Numerical solutions for the functions 
of the $\kappa=+1/2$ fermion zero mode (\ref{zmplushalf}) on the symmetry axis, 
with $SU(3)$ background gauge field given
by Eqs.~(\ref{Asphal-like}) and (\ref{fnum})
for $\zeta \equiv r/(r+1)$ and  $r=|z|$.
Left panel: $l_6^{(+)}(\zeta)$ and $-l_1^{(+)}(\zeta)$ 
are shown as broken and solid curves, respectively.
Right panel: $l_2^{(+)}(\zeta)$ and $l_4^{(+)}(\zeta)$ 
are shown as broken and solid curves, respectively
[not shown are $l_3^{(+)}(\zeta)$ and $l_5^{(+)}(\zeta)$, 
which take values close to zero].} 
\label{FIG-ZMplus} 
\end{figure}

In order to be specific, we take the following radial gauge field function:
\begin{equation}
f(\zeta)= 
\left\{
\begin{array}{ll}
0   & \quad\mathrm{if}\;\;\zeta \leq 1/4 \\
(4\,\zeta-1)^2 / \left[(4\,\zeta-1)^2+(4\,\zeta-3)^2\right]
 &  \quad\mathrm{if}\;\;\zeta \in  (1/4,3/4) \\
1   & \quad\mathrm{if}\;\; \zeta \geq 3/4 
\label{fnum}
\end{array}
\right. \,,
\end{equation}
in terms of the compact coordinate $\zeta \in [0,1]$ defined by
\begin{equation}
\zeta \equiv r/(r+1)\,.
\label{zeta}
\end{equation}
The numerical solutions for the  
functions of the $\kappa=-1/2$ and $\kappa=+1/2$
zero modes on the axis
are shown in Figs.~\ref{FIG-ZMminus} and \ref{FIG-ZMplus}, respectively.
A complete numerical solution of these zero modes over the 
$(r,\theta)$--halfplane is left to a future publication.

The analysis of this section
does not rigorously prove the existence of fermion
zero modes. It is, in principle, possible that the only solution of
the zeromode equation (\ref{zeromodeeqn}) which matches the asymptotic
behavior (\ref{Psihat-infinity}) is
the trivial solution vanishing everywhere. But the existence of two
normalizable solutions on the symmetry axis, which is already a 
nontrivial result, suggests that the  partial
differential equation (\ref{zeromodeeqn}) has indeed two normalizable
solutions. 
In fact, the symmetries of the $\widehat{\mathrm{S}}$
\emph{Ansatz} may play a crucial role for the existence 
of fermion zero modes, just as was the case for the $\mathrm{S}$
and $\mathrm{S}^\star$ sphalerons.

\section{Conclusion}
\label{sec:Conclusion}

As explained in the Introduction and in more detail in 
Ref.~\cite{NA85}, one pair of fermion zero modes 
(with a cone-like pattern of level crossings) 
is needed to recover the $SU(3)$ Bardeen anomaly as a
Berry phase (Fig.~\ref{FIG-NCSbardeen}). 
The existence of one pair of fermion zero modes somewhere 
inside the basic \ncs~of gauge-transformed vacua 
follows from the family index theorem. In this article, we 
have argued that these zero modes are, most likely, related to
a new sphaleron solution of $SU(3)$ \YMHth, 
indicated by $\widehat{\mathrm{S}}$ in Fig.~\ref{FIG-SSstarSbar}.

The self-consistent  \emph{Ans\"{a}tze} for the bosonic and fermionic 
fields of $\widehat{\mathrm{S}}$ 
have been presented in Secs. \ref{sec:Bosonic fields} 
and \ref{sec:Ansatzandzeromodeequation}. 
Specifically, the gauge and Higgs fields are given by
Eqs.~(\ref{AsphericalAnsatz})--(\ref{alphajparity}) and 
(\ref{PhiAnsatz})--(\ref{betakparity}), and the fermionic 
field by Eq.~(\ref{K3eigenstate}).
It appears that the reduced field equations can only be solved numerically. 
The sphaleron $\widehat{\mathrm{S}}$  has at least 
three negative modes, if it is indeed at the ``top'' of a 
noncontractible three-sphere in configuration space [cf. the discussion 
in the paragraph starting a few lines below Eq.~(\ref{symmetriesVW})].

In pure $SU(3)$ \YMth, there exists perhaps a corresponding 
Euclidean solution, just as the BPST instanton
$\mathrm{I}$ \cite{BPST75} corresponds to the sphaleron $\mathrm{S}$ 
(the gauge field of a three-dimensional
slice through the center of $\mathrm{I}$ roughly matches that
of $\mathrm{S}$).
This non-self-dual solution would have at least two negative modes. 
It is, however, not guaranteed that the reduced field equations
have a localized solution in $\bR^4$.
For a brief discussion of non-self-dual solutions, 
see, e.g., Section~6 of Ref.~\cite{K93npb} and references therein.

Purely theoretically, it is of interest to have found the 
sphaleron $\widehat{\mathrm{S}}$ in $SU(3)$ \YMHth~which 
may be ``responsible'' for the non-Abelian (Bardeen) 
anomaly. More phenomenologically, there is the possibility that 
$\widehat{\mathrm{S}}$--like 
configurations take part in the dynamics of Quantum 
Chromodynamics (QCD) or even grand-unified theories. 
The $SU(3)$ Bardeen anomaly cancels, of course, between 
the left- and right-handed quarks of QCD, but the configuration space of 
the gluon gauge field still has nontrivial topology. Quantum effects 
may then balance the $SU(3$) gauge field configuration given by 
Eqs. (\ref{AsphericalAnsatz})--(\ref{alphajparity}). The resulting 
effective sphaleron ($\widehat{\mathrm{S}}_\mathrm{QCD}$) can 
be expected to play 
a role in the nonperturbative dynamics of the strong interactions.

\begin{appendix}
\section{Ansatz energy density}
\label{sec:AppendixA}

In this appendix, the energy density from the
\emph{Ansatz} gauge and Higgs fields of 
Eqs.~(\ref{AsphericalAnsatz}) and (\ref{PhiAnsatz}) is given.
Specifically, the contributions to the energy density  (\ref{edens}) 
from the Yang--Mills term, the kinetic Higgs term, and the Higgs potential 
term in the energy (\ref{energyfunctional}) are
\mathindent=0cm
\begin{align}
&\widehat{e}_{\rm \,YM}= \;
\frac{1}{2\,g^2\,r^2\sin^2\theta} \, \Bigl\{
\cos^2\theta\, \bigl[\, 
 \partial_r \alpha_1 +
\left( \alpha_9 + \alpha_4\alpha_9-\half
    \alpha_2\alpha_{11}-\half  \alpha_3\alpha_{10}
  \right)/r
\,\bigr]^2 \no \\[1mm]
& + \bigl[\,
 \partial_r \alpha_2 - \bigl(
  \alpha_{10} -\half
  \alpha_4\alpha_{10}  
+\half \cos^2\theta\, \alpha_3\alpha_9 - \half \sqrt{3}\,
    \alpha_5\alpha_{10} -\half \cos^2\theta\, \alpha_1 \alpha_{11}
  \bigr)/r
\,\bigr]^2 \no\\[1mm]
& + \cos^2\theta\, \bigl[\,
 \partial_r \alpha_3 -
\left( 2\alpha_{11} - \half \alpha_2\alpha_9 +\half
    \alpha_4\alpha_{11} -\half \sqrt{3}\, \alpha_5\alpha_{11} - \half
     \alpha_1\alpha_{10} \right)/r \,\bigr]^2 \no\\[1mm]
& + \bigl[\,
 \partial_r \alpha_4 + 
\left( \half \cos^2\theta \,\alpha_3\alpha_{11} - \cos^2\theta\,
  \alpha_1\alpha_9 
  -\half \alpha_2\alpha_{10} \right)/r
\,\bigr]^2 \no\\[1mm]
& + \bigl[\,
\partial_r\alpha_5 -
\half \sqrt{3} \left( \alpha_2
  \alpha_{10} + \cos^2\theta\, \alpha_3\alpha_{11} \right)/r
\,\bigr]^2
\,\Bigr\} \no
\end{align}
\begin{align}
& +\frac{1}{2\,g^2\,r^2} \, \Bigl\{
\bigl[\,
 \partial_r \alpha_6
     -\left( \cos\theta\,\partial_\theta \alpha_9 - \half
       \alpha_8\alpha_{10} +\half \cos^2\theta\, \alpha_7 \alpha_{11}
       - \sin\theta\, \alpha_9 \right)/r
\,\bigr]^2 \no \\[1mm]
& + \bigl[\,
 \cos\theta\, \partial_r \alpha_7 - \left(
  \partial_\theta \alpha_{10} - \half \cos\theta \,\alpha_6
  \alpha_{11} +\half\cos\theta\, \alpha_8\alpha_9 \right)/r
\,\bigr]^2 \no \\[1mm]
& + \bigl[\,
 \partial_r \alpha_8 - \left(
  \cos\theta\, \partial_\theta\alpha_{11} +\half \alpha_6\,\alpha_{10}
  - \half \cos^2\theta\, \alpha_7\alpha_9 -\sin\theta\, \alpha_{11} 
  \right)/r \,\bigr]^2 \,\Bigr\}  \no 
\\[1mm]
%\end{align}
%\begin{align}
& + \frac{1}{2\,g^2\,r^4\sin^2\theta} \, \Bigl\{
\bigl[\,
 \alpha_6 -\sin\theta\, \alpha_1 
-\half \alpha_2\alpha_8 
+\alpha_4\alpha_6  -\half \cos^2\theta \,\alpha_3\alpha_7 
+\cos\theta\,\partial_\theta \alpha_1
\,\bigr]^2 \no \\[1mm]
& + \bigl[\,
 \cos\theta\, \alpha_7 +\half 
\cos\theta\, \left(\alpha_3\alpha_6 - \alpha_1\alpha_8
-\sqrt{3}\,\alpha_5\alpha_7 - \alpha_4\alpha_7\right) -
\partial_\theta \alpha_2
\,\bigr]^2 \no \\[1mm]
& + \bigl[\,
 2\alpha_8 +\sin\theta\,\alpha_3 +
\half 
\alpha_4\alpha_8-\half \alpha_2\alpha_6-\half \sqrt{3}\,
\alpha_5\alpha_8 
-\half \cos^2\theta\,
\alpha_1\alpha_7  -\cos\theta\, \partial_\theta \alpha_3
\,\bigr]^2 \no \\[1mm]
& + \bigl[\,
 \cos\theta
\left( \alpha_1\alpha_6 + \half \alpha_2\alpha_7-\half
  \alpha_3\alpha_8 \right)-  \partial_\theta \alpha_4
\,\bigr]^2 \no \\[1mm]
& + \bigl[\,
\half \sqrt{3}\cos\theta\left(
  \alpha_3\alpha_8+\alpha_2\alpha_7\right) -
\partial_\theta \alpha_5
\,\bigr]^2 
\,\Bigr\}\,,
\label{energydensityYM}
\end{align}

\mathindent=0cm
\begin{align}
&\widehat{e}_{\rm \,Hkin} =\;\eta^2\, \Bigl\{ 
\bigl[\,
\partial_r \beta_1 -\half
    \left(  \cos^2\theta\, \alpha_9\beta_2 +  \alpha_{10} 
    \beta_3 \right)/r
\,\bigr]^2 + \cos^2\theta
\no\\[1mm]& 
\times \bigl[\,
 \partial_r\beta_2 +\half \, \left(
    \alpha_9\beta_1 - \alpha_{11} \beta_3 \right)/r
\,\bigr]^2 
+\bigl[\,
 \partial_r\beta_3 +\half\,
    \left(\,  \alpha_{10}\beta_1 + 
    \cos^2\theta\, \alpha_{11} \beta_2 \right)/r
\,\bigr]^2
\Bigl\} \no
\\[1mm]
%\end{align}
%\begin{align}
& + \frac{\eta^2}{r^2}\, \Bigl\{
\bigl[\,
 \partial_\theta \beta_1 - \half
\cos\theta\, \left( \alpha_7\beta_3 + \alpha_6\beta_2\right)
\,\bigr]^2 
+\bigl[\,
 \partial_\theta \beta_3 + \half \cos\theta\, \left(
\alpha_8\beta_2 + \alpha_7\beta_1 \right) \,\bigr]^2 
\no\\[1mm]
& +\bigl[\,
  \cos\theta \, \partial_\theta \beta_2 -\sin\theta\,\beta_2 
  + \half \left( \alpha_6\beta_1 - \alpha_8 \beta_3 \right)
\,\bigr]^2 
\,\Bigr\} \no 
\\[1mm]
%\end{align}
%\begin{align}
& + \frac{\eta^2}{r^2\sin^2\theta} \Bigl\{
 \quart\, \bigl[\,
\alpha_4\beta_1 + (1/\sqrt{3}\,)\,\alpha_5\beta_1 + 
  \cos^2\theta \, \alpha_1 \beta_2 +   \alpha_2\beta_3
\,\bigr]^2 \no\\[1mm]
& + \quart\,\cos^2\theta\,\bigl[\,
2 \beta_2 - \alpha_1\beta_1 +  \alpha_4\beta_2 
-(1/\sqrt{3}\,)\, \alpha_5 \beta_2 -  \alpha_3\beta_3
\,\bigr]^2 \no\\[1mm]
& + \quart\, \bigl[\,
2 \beta_3 +\alpha_2\beta_1 -(2/\sqrt{3}\,)\, \alpha_5\beta_3 + 
  \cos^2\theta \, \alpha_3 \beta_2
\,\bigr]^2 
\,\Bigr\}\,,
\label{energydensityHkin}
\\[2ex]
&\widehat{e}_{\rm \,Hpot} =\;
\lambda \,\eta^4 \,
\bigl[\, \beta_1^2+ \cos^2\theta\, \beta_2^2 + \beta_3^2 -1 \,\bigr]^2
  \,.
\label{energydensityHpot}
\end{align}
\mathindent=2em 
These energy densities are manifestly symmetric under reflection
$\theta \to \pi - \theta$, provided the \emph{Ansatz} functions 
$\alpha_j(r,\theta)$ and $\beta_k(r,\theta)$
are; cf. Eqs.~(\ref{alphajparity}) and (\ref{betakparity}).

\section{Argument for a nontrivial regular solution}
\label{sec:AppendixB}  

In this appendix, we  present a heuristic argument for the existence 
of a nontrivial regular solution of the reduced bosonic 
field equations, with 
\emph{Ansatz} functions $\alpha_j(r,\theta)$ and $\beta_k(r,\theta)$ 
from Eqs.~(\ref{AsphericalAnsatz})--(\ref{alphajparity}) and 
(\ref{PhiAnsatz})--(\ref{betakparity}).
The reasoning here complements the one of Section~\ref{sec:Energy}.
It focuses on different energy contributions  
from the region near the equatorial plane [$\theta=\pi/2$ or $z=0$]
and uses a simple scaling argument to show that the total energy of a regular 
solution cannot be zero. The coupling constant $\lambda/g^2$
is taken to be nonzero. 

First, consider the possibility that $\beta_1(r,\theta)$ has a vacuum value 
at the coordinate origin, $\beta_1(0,\pi/2) = \pm\, 1$. 
We now try to get a vacuum solution over the whole $(r,\theta)$--halfplane 
but do not allow for singular configurations with a finite change of  
$\beta_1(0,\pi/2)$ over an infinitesimal $r$ interval at the origin $r=0$.
A vanishing Higgs potential energy density (\ref{energydensityHpot}) 
on the equatorial plane then requires 
$\beta_3(r,\pi/2)=-[\,1-\beta_1(r,\pi/2)^2\,]^{1/2}$, 
assuming $\beta_1(r,\pi/2)$ to change monotonically from the  
value $\pm\, 1$ at $r=0$ to the asymptotic value $0$ 
as given by the boundary condition (\ref{betaBCSinfinity}). 
Define the length scale $R$ by $\beta_3(R,\pi/2) =-1/2\,$.

With $\beta_1^2 + \beta_3^2 \approx 1$ near the equatorial plane, 
a vanishing first curly bracket in the 
kinetic Higgs energy density (\ref{energydensityHkin})
requires $\alpha_{10} \approx -2\, (r\, \partial_r \beta_3)/\beta_1
\approx 2\, (r\, \partial_r \beta_1)/\beta_3$, 
which peaks somewhere around $r=R$. The corresponding contribution
of $\alpha_{10}$ to the \YM~energy typically scales as $1/R$. 
This energy contribution can be reduced by increasing the 
value of $R$. But for large $R$ and $\beta_k(r,\theta)$
of the asymptotic form (\ref{betasfromW}),
the $\theta$--dependence of  $-2\, (r\, \partial_r \beta_3)/\beta_1$ 
and $2\, (r\, \partial_r \beta_1)/\beta_3$ is entirely different.
Hence, the previous expression for $\alpha_{10}(r,\theta)$  
needs to be modified near $\theta \approx \pi/2$ for large $r$ and
the kinetic Higgs energy term picks up  
a contribution which typically scales as $R$.
Clearly, both terms cannot be reduced to zero simultaneously
by changing $R$. This basically rules out having a regular vacuum 
solution with $|\beta_1(0,\pi/2)| = 1$.

Second, consider the possibility that the value of $|\beta_1(0,\pi/2)|$
does not equal $1$. 
Since  $\beta_2(r,\theta)$ and  $\beta_3(r,\theta)$ vanish on the
whole $z$--axis by the boundary conditions (\ref{betaBCSaxis}), 
one then has for the Higgs field  (\ref{PhiAnsatz})
at the origin $|\widehat\Phi(0,0,0)|\ne \eta$, which already 
differs from the vacuum solution with $|\Phi(r,\theta,\phi)|=\eta$ everywhere. 

Apparently, both possibilities lead to 
a regular solution with nonzero total energy,
which concludes our heuristic argument. The actual 
solution may very well have a Higgs field $\widehat\Phi(r,\theta,\phi)$ 
that vanishes at the coordinate origin, with a corresponding 
localized nonzero energy density; cf. Section ~\ref{sec:Energy}.

\section{Fermion zero modes from gauge fields with winding number \boldmath$-1$}
\label{sec:AppendixC}

Instead of the generator $U$ of 
$\pi_5[SU(3)]$ 
with winding
number $+1$ [given by Eq.\ (\ref{Udef})], 
one may use a generator $\widetilde U$ with winding number
$-1$. This matrix $\widetilde U$ is simply the inverse of $U$,
namely $\widetilde U(\psi,\mu,\alpha,\theta,\phi)
=U^{-1}(\psi,\mu,\alpha,\theta,\phi)$.
The background gauge field for the fermion zero modes is then
determined by the matrix function $\widetilde
W(\theta,\phi)\equiv W^{-1}(\theta,\phi)$, with 
$g \widetilde A_\mu=- f(r)\, \partial_\mu \widetilde W\, \widetilde W^{-1}$
for the approximate anti-sphaleron.

The corresponding Dirac Hamiltonian 
$\widetilde H \equiv \i\, \sigma_m \,(\partial_m-g \widetilde A_m)$ 
has the symmetry properties
\begin{equation}
[\widetilde K_3, \widetilde H]=0\,,\quad
\{R_3,\widetilde H\}=0\,,
\end{equation}
where
\begin{align}
\widetilde K_3 &\equiv
 -\i \pr_\phi -\frachalf \left(  \lambda_3-\sqrt{3}\,\lambda_8\right)
+ \frachalf \sigma_3\,,
\end{align}
and $R_3$ is still given by (\ref{R3def}).
An eigenstate of $\widetilde K_3$ with eigenvalue $\kappa$ has the form
\begin{equation}
\widetilde \Psi(r,\theta,\phi) = \vect{
\widetilde F_{R+}(r,\theta)\,\, \e^{\i\, (\kappa-1/2)\, \phi}\\[1ex]
\widetilde F_{G+}(r,\theta)\,\, \e^{\i\, (\kappa-3/2)\, \phi}\\[1ex]
\widetilde F_{B+}(r,\theta)\,\, \e^{\i\, (\kappa+1/2)\, \phi}\\[1ex]
\widetilde F_{R-}(r,\theta)\,\, \e^{\i\, (\kappa+1/2)\, \phi}\\[1ex]
\widetilde F_{G-}(r,\theta)\,\, \e^{\i\, (\kappa-1/2)\, \phi}\\[1ex]
\widetilde F_{B-}(r,\theta)\,\, \e^{\i\, (\kappa+3/2)\, \phi}} \,.
\end{equation}

The zeromode equation $\widetilde H\, \widetilde\Psi=0$ now
reduces to
\begin{equation}
\left\{ A_1(\theta)\, \pdq{}{r} + \frac{1}{r} \, A_2(\theta)\,
  \pdq{}{\theta} + \frac{1}{r}\, \widetilde A_3(\kappa)\, + \frac{f(r)}{r} \,
  \widetilde A_4(\theta)
  \right\} \, \widetilde \Psi(r,\theta,0) =0\,.
\label{zeromodeeqninverse}
\end{equation}
Here, the matrices $A_1$ and  $A_2$ are given by  
Eqs.~(\ref{A1def}) and (\ref{A2def}), the matrices
$\widetilde A_3$ and $\widetilde A_4$ by
\begin{equation}
\widetilde A_3(\kappa) = N\, A_3(-\kappa)\, N^{-1}\,,\quad
\widetilde A_4(\theta) = N \, A_4(\pi-\theta)\, N^{-1}\,,
\end{equation}
with $A_3$ and $A_4$ from Eqs.~(\ref{A3def}) and (\ref{A4def}), 
and the constant matrix $N$ by
\begin{equation}
N \equiv 
\twomat{0_3 & \threematw{1 & 0 & 0 \\[-1ex] 0 & -1 & 0 \\[-1ex] 0 & 0 & -1}\\[-1ex]
\threematw{1 & 0 & 0 \\[-1ex] 0 & -1 & 0 \\[-1ex] 0 & 0 & -1} & 0_3}\,.
\end{equation}
Note that the matrices $A_1$ and $A_2$ have the following reflection
properties:
\begin{equation}
A_1(\theta) = N\,A_1(\pi-\theta)\, N^{-1} \,, \quad
A_2(\theta) = - N\, A_2(\pi-\theta)\, N^{-1}\,.
\end{equation}
Hence, there is a one-to-one correspondence between the solutions of
Eqs.~(\ref{zeromodeeqn}) and (\ref{zeromodeeqninverse}): 
$\widehat\Psi(r,\theta,\phi)$ solves (\ref{zeromodeeqn}) and has
$K_3$--eigenvalue $\kappa$
if and only if
$\widetilde\Psi(r,\theta,\phi) \equiv
N \widehat \Psi(r, \pi-\theta,\phi)$ solves
(\ref{zeromodeeqninverse}) and has $\widetilde K_3$--eigenvalue $-\kappa$.

\end{appendix}

\end{document}